\documentclass{caosp310}

\usepackage{graphicx}
\usepackage{adjustbox}
\usepackage{natbib}
\usepackage[hyphens]{xurl}
\usepackage{rotating}
\usepackage{float}
\usepackage{xcolor}
\usepackage{listings}

\definecolor{mygreen}{rgb}{0.15,0.55,0.4}
\definecolor{mymagenta}{rgb}{0.58,0,0.82}
\definecolor{myred}{rgb}{1, 0.6, 0}
\definecolor{myblue}{rgb}{0.2, 0.4, 0.9}
\definecolor{mygrey}{rgb}{0.67, 0.67, 0.67}

\lstdefinestyle{myListingStyle}{
    backgroundcolor=\color{white},   
    basicstyle=\ttfamily\footnotesize, 
    breaklines=true,                   
    frame=single,                      
    captionpos=b,                      
    commentstyle=\color{mygrey}, 
    morecomment=[l]{\#}, 
    stringstyle=\color{mymagenta},      
    keywordstyle=\color{myred}\bfseries, 
    keywordstyle=[2]{\color{myblue}},
    keywordstyle=[3]{\color{orange}\bfseries},
    morekeywords={confirm, find, lightcurve, crossmatch, monitor, classify, notify, store, publish}, 
    morekeywords=[2]{object_id, broker, brokers, source, sources, filters, required, catalog, stream, method}, 
    morekeywords=[3]{all, },              
    numbers=left, 
    numberstyle=\tiny, 
    stringstyle=\color{mygreen}\bfseries,
    morestring=[b]", 
}

\lstdefinestyle{xxx}{
    basicstyle=\small\ttfamily, 
    breaklines=true, 
    numbers=left, 
    numberstyle=\tiny, 
    frame=single, 
    language=DSL 
}

\articleNo{?}
\pubyear{2026}
\volume{56}
\volnumber{1}
\firstpage{176}
\lastpage{183}
\received{September 25, 2025}
\accepted{November 19, 2025}

\def\BibTeX{{\rm B\kern-.05em{\sc i\kern-.025em b}\kern-.08em
             T\kern-.1667em\lower.7ex\hbox{E}\kern-.125emX}}

\begin{document}

\hauthor{V.\,Vuj{\v c}i{\' c}, V.A.\,Sre{\' c}kovi{\' c}, S.\,Babarogi{\' c}}

\title{Alertissimo - a tool for orchestration of LSST broker streams}

\author{
               V.\,Vuj{\v c}i{\' c}\inst{1}\orcid{0000-0002-0525-1197}
      \and
        V.A.\,Sre{\' c}kovi{\' c}\inst{2}\orcid{0000-0001-7938-5748}
      \and
        S.\,Babarogi{\' c}\inst{3}\orcid{0000-0003-3635-4671}
}

\institute{
            Astronomical Observatory, Volgina 7, 11060 Belgrade, Serbia \email{veljko@aob.rs}
         \and 
                      University of Belgrade, Institute of Physics Belgrade, PO Box 57, 11001 Belgrade, Serbia
         \and 
           University of Belgrade, Faculty of Organizational Sciences, Jove Ili{\' c}a, 11000 Belgrade, Serbia
          }

\date{March 8, 2003}

%
%

\maketitle
\begin{abstract}
The Vera C. Rubin Observatory, through its Legacy Survey of Space and Time, will soon start producing 10 million alerts on transient astronomical objects per night. Due to logistics and bandwidth, alerts will not be dispatched directly to the public but to 'brokers' i.e. tools selected by LSST to handle alert streams. Brokers offer both common, specific and micro-specific functionalities related to alert handling, analysis, representation and dissemination. In this ecosystem, potentially augmented by data streams from other astronomical sources, there is a - need demonstrated by the community - for use cases which combine features of individual brokers. In this paper we present initial efforts and a prototype of such a tool, along with a language that would allow users to define use cases / workflows in a manner tailored for the domain.
\end{abstract}

\section{Introduction}
\label{intr}

As a new era of astronomical data science is about to kick off with Vera C. Rubin Observatory's Legacy Survey of Space and Time (Rubin/LSST) \citet{ive19} production phase, the ecosystem of tools and infrastructure dealing with transient astronomical events looks fit and (almost) ready for the data burst. It is commonly known that the LSST Alert Production pipeline \citep{bosch2018overview} will capture and process images, perform difference-image analysis and distribute alerts of every astronomical source with a signal-to-noise ratio (SNR) ratio $>$ 5 in positive or negative flux \citep{graham2019lsst}. Due to bandwidth and other considerations, alerts will not be directly available to the public but through 'brokers' - tools developed by different scientific teams worldwide, specifically made for and approved by LSST. At least six brokers (AleRCE, AMPEL, ANTARES, Fink, Lasair, Pitt-Google) will be able to ingest LSST alert stream in near-real time, analyze, classify, offer UI tools, programming APIs and more. Some of the main features of brokers overlap but they differ significantly in how they expose these features, with even greater divergence in their underlying implementations and backend technologies. \citep{vujcic2025overview}.   

\section{Alertissimo}
\label{bro}

We introduce \cite{ale25} - a tool for orchestration of broker streams and potentially for a wider scope of astro-science use cases. It can be seen as a novel topic but also an extension of the work done and analyzed by the Serbian group related to astronomical stream processing and virtual observatory \citep{vujcic2020real,jevremovic2020databases}. This tool is conceived on two premises: \\ a) that there is a need for building workflows out of multiple LSST broker data streams (and potentially external sources as well); \\ b) that there is a need for a tool that would allow expressing such workflows in a powerful yet approachable manner. \\ We can roughly name these two premises as 'Broker Orchestration' and 'DSL for Transients'. We also present Alertissimo's overall architecture. 
\subsection{Broker orchestration}
As stated above, at least 6 LSST brokers are in a mature development phase, and some of them - like Lasair, ALeRCE, ANTARES and Fink - have worked in production with Zwicky Transient Facility (ZTF) data (\citealt{ztf2019}, data volume order of magnitude lower than LSST). Apart from technological choices, implementation differences and performance, brokers also vary in features (and microfeatures), and in the ways they are exposed through their UIs and APIs. There are ongoing discussions in the Transients and Variable Stars scientific collaboration (LSST-TVS) on use cases that combine features of various brokers (see the Appendix \ref{appendix} for a use case featuring supermassive binary black hole (SMBBH) detection). There may also be availability and responsiveness considerations, in a technical sense. 

Brokers offer external access to their features through APIs, most through REST\footnote{Representational State Transfer, an architectural style for web services which defines GET and POST methods for data retrieval} and Python, some through a Python interface only. Here we have a similar but non-standardized set of their exposed methods, not only related to major feature differences but also to naming conventions and microfeatures (e.g - how do they retrieve single vs multiple objects? can they handle SQL? how does output look like and can you control granulation? etc). API access control is implemented differently across brokers, typically requiring users to supply credentials — such as tokens or username/password combinations — with varying permission models and granularity.

It is a relatively easy task to write a program or a script that would pick specific methods from various brokers and perform a single use case. Alertissimo should act here as a generator of scientific scenarios - take any number of features from any broker (or other available source) and orchestrate them with one another. The primary function of Alertissimo here is one of a science-enabler, where all decisions are in control of the user\footnote{However, the idea persists that Alertissimo could also act in a 'light' AI-based counseling manner too, based on the knowledge of previous user experiences/success rates. This kind of development could take place in later phases.}.

Alertissimo is designed with an input-agnostic core, where user-specified workflows are first translated into intermediate representations (IR) with a well-defined structure built using Pydantic\footnote{Pydantic is a Python library designed for data validation, parsing, and serialization which provides a structured way to define data schemas and ensures that incoming data conform to the specified types and constraints.} models. These IR models are the product of a conceptual analysis of broker features; their design is generalized, while their implementation is specialized through a polymorphic class structure.

\begin{figure}
 \begin{center}
 \includegraphics[width=0.70\columnwidth]{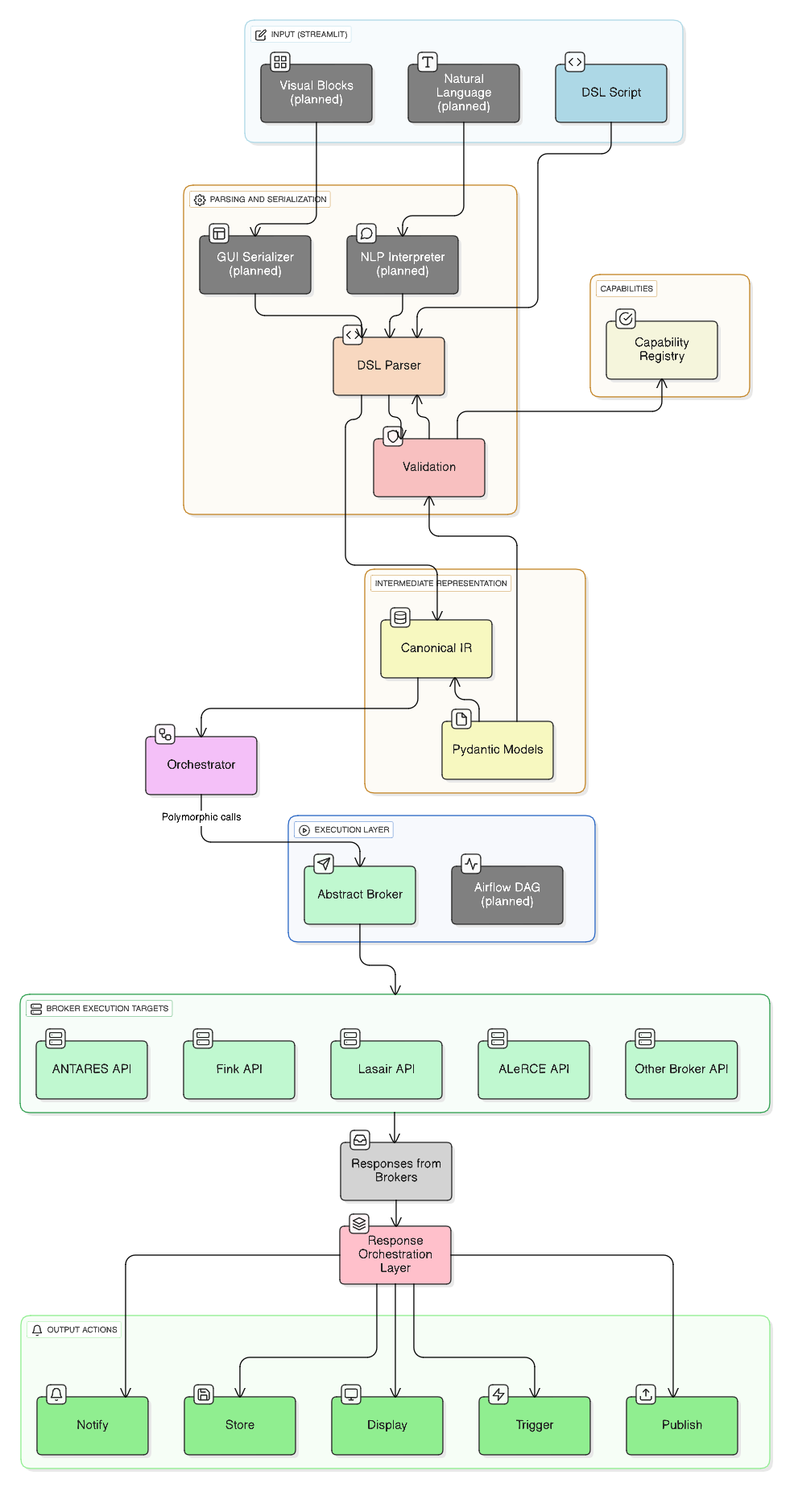}
  \caption{End to end flow diagram of Alertissimo modules illustrating the pipeline from DSL input through orchestration to broker execution. For detailed description see Subsection \ref{architecture}}
 \label{fig:flowchart}
 \end{center}
\end{figure}

\subsection{DSL for transients}
The challenge of translating high-level user specifications into executable workflows is often addressed through Domain-Specific Languages (DSLs) and workflow systems \citep{gil2010wings}. There are three main directions for Alertissimo UI development, one of them being the backbone of the others - the DSL for Transients. The two others, Natural Language Processing (NLP) \citet{chowdhary2020natural} and visual workflow builder are planned to be built once a mature DSL foundation is established. DSLs \citep{mernik2005} are programming languages designed for a very specific purpose and are usually of declarative nature\footnote{Declarative programming languages are stating 'what' should be done in contrast with imperative languages which are describing 'how' it should be done. All general programming languages are imperative while more narrowly specified languages, such as SQL, rule-based languages, HTML etc are declarative.}.  
Alertissimo's DSL for Transients acts as the semantic and syntactic backbone linking user intent with system capabilities. It exposes a structured, declarative representation of broker operations, allowing every workflow to be expressed in a consistent formalism. This provides that all interactions, no matter how informal their starting point was, ultimately resolve into a precise and testable specification. The demo version of Alertissimo featuring DSL for Transients is currently available through a github branch (\url{https://github.com/sambolino/alertissimo/tree/feature/dsl}). Example of DSL for Transients is found in the Appendix \ref{appendix}.

The natural language interface in Alertissimo will enable users to describe their scientific intentions in plain terms, while maintaining expressive power. The aim is to test both a pure NLP prompt, where users would describe use-cases in written English, and a chat-bot style AI dialogue which could act as an assistant providing refinement and disambiguation, even for users initially unfamiliar with broker architectures. Both the natural language and visual interfaces of Alertissimo will be developed on top of a formally defined DSL grammar. As DSL defines the valid constructs, relationships, and parameter types for workflows involving broker orchestration, it also ensures that any interface — textual, conversational, or visual — ultimately produces valid, executable specifications. This formal grammar serves as the common translation layer: natural language inputs are parsed and normalized into DSL expressions, and also the visual interface directly maps user manipulations of blocks and links into equivalent DSL statements.

From a broader perspective, this layered UI approach reflects a continuum between accessibility and expressiveness. At one end, the conversational interface empowers domain scientists to describe ideas without syntactic constraints; at the other, the DSL offers power users fine-grained control and versionable scripts suitable for integration into research pipelines. The visual interface bridges these modes, exposing the structure of the DSL while remaining approachable. 

\subsection{Architecture overview}
\label{architecture}

The overall architecture of Alertissimo is depicted in Figure \ref{fig:flowchart}, with data flow and key components as follows:

\begin{enumerate}
    \item Input Layer: Users can define their use-cases via multiple interfaces (DSL, NLP/chatbot (planned), Visual blocks - Yahoo Pipes \citep{pruett2007yahoo} style (planned)), with the DSL being the currently implemented method.
    \item Parsing and Validation: User input is reduced to DSL formalism to the appropriate parser (DSL Parser, etc.). Validation is performed both in terms of grammar and broker capabilities.
    \item Capabilities: Broker capabilities are defined within yaml\footnote{YAML is a human-readable data serialization language primarily used for configuration files and data storage} files
    \item IR Generation: Valid DSL is being translated into IR models, still being execution-agnostic.
    \item Orchestration and Execution: The validated IR is processed by the Orchestrator. This component manages the workflow by making polymorphic calls to the Execution Layer, which consists of an Abstract Broker interface. This design allows the system to specialize the implementation for different targets while maintaining a generic core. For the sake of performance/execution control, The orchestrator can also trigger a workflow coordination framework, such as an Airflow DAG (planned).
    \item Response Handling and Output: Responses from brokers are collected and processed by and the system executes one or more output actions, such as to Notify a user, Store the data, Display it, Trigger a downstream process, or Publish the results.
\end{enumerate}

\section{Conclusion}
\label{con}

Alertissimo is a new tool for building scientific workflows out of streams of alerts generated by Rubin/LSST. The alerts are not disseminated directly to the public, but through tools called 'brokers' - projects developed for ingestion and analysis of LSST transient alerts. Alertissimo covers all of the brokers' concepts and features and offers an integrated interface for orchestrating multiple brokers within a single workflow. On top of Alertissimo lies a domain-specific language devised to easily yet expressively represent scientific use cases. NLP/LLM/chatbot and visual workflows are planned as extensions that dependably translate and validate to the canonical DSL. 

Together, these interaction paradigms establish a scalable human–machine interface strategy, ensuring that Alertissimo remains adaptable as new brokers, data modalities, and analysis paradigms emerge in astronomy and astrophysics. Our prototype demonstrates how broker interoperability and workflow definition can empower the astronomy and astrophysics community in the era of Rubin/LSST.


%
%

\acknowledgements
This research was supported by the Ministry of Science, Technological Development and Innovation of the Republic of Serbia (MSTDIRS) through contract no. 451-03-66/2024-03/200002 made with Astronomical Observatory (Belgrade), 451-03-47/2023-01/200024 made with Institute of physics Belgrade.
The authors acknowledge the networking opportunities from the COST Action CA22133 - The birth of solar systems (PLANETS) supported by COST (European Cooperation in Science and Technology).

\bibliography{Caosp_LSST}

@article{vujcic2025overview,
  title={An overview of astronomical transient brokers in Rubin era},
  author={Vuj{\v{c}}i{\'c}, V and Sre{\'c}kovi{\'c}, VA and Babarogi{\'c}, S and Aleksi{\'c}, J},
  journal={Contrib. Astron. Obs. Skalnat{\'e} Pleso},
  volume={55},
  number={2},
  pages={95--105},
  year={2025}
}

@article{bosch2018overview,
  title={An overview of the LSST image processing pipelines},
  author={Bosch, James and AlSayyad, Yusra and Armstrong, Robert and Bellm, Eric and Chiang, Hsin-Fang and Eggl, Siegfried and Findeisen, Krzysztof and Fisher-Levine, Merlin and Guy, Leanne P and Guyonnet, Augustin and others},
  journal={arXiv preprint arXiv:1812.03248},
  year={2018}
}

@book{graham2019lsst,
  title={LSST Alerts: Key Numbers},
  author={Graham, ML and Bellm, E and Guy, L and Slater, CT and Dubois-Felsmann, G},
  year={2019},
url={https://github.com/lsst-dm/dmtn-102},
  publisher={DMTN-102, URL https://dmtn-102. lsst. io, LSST Data Management Technical Note}
}

@incollection{vujcic2020real,
  title={Real-time stream processing in astronomy},
  author={Vuj{\v{c}}i{\'c}, Veljko and Jevremovi{\'c}, Darko},
  booktitle={Knowledge Discovery in Big Data from Astronomy and Earth Observation},
  pages={173--182},
  year={2020},
  publisher={Elsevier}
}

@article{jevremovic2020databases,
  title={Databases for collisional and radiative processes in small molecules needed for spectroscopy use in astrophysics},
  author={Jevremovi{\'c}, D and Sre{\'c}kovi{\'c}, VA and Marinkovi{\'c}, BP and Vuj{\v{c}}i{\'c}, V},
  journal={Contributions of the Astronomical Observatory Skalnate Pleso},
  volume={50},
  number={1},
  pages={44--54},
  year={2020}
}

@article{mernik2005,
  title={When and how to develop domain-specific languages},
  author={Mernik, Marjan and Heering, Jan and Sloane, Anthony M},
  journal={ACM computing surveys (CSUR)},
  volume={37},
  number={4},
  pages={316--344},
  year={2005},
  publisher={ACM New York, NY, USA}
}

@article{chowdhary2020natural,
  title={Natural language processing},
  author={Chowdhary, KR1442},
  journal={Fundamentals of artificial intelligence},
  pages={603--649},
  year={2020},
  publisher={Springer}
}

@article{komossa2025extremes,
  title={The extremes of AGN variability: outbursts, deep fades, changing looks, exceptional spectral states, and semi-periodicities},
  author={Komossa, S and Grupe, D and Marziani, P and Popovi{\'c}, L{\v{C}} and Mar{\v{c}}eta-Mandi{\'c}, S and Bon, E and Ili{\'c}, D and Kova{\v{c}}evi{\'c}, AB and Kraus, A and Haiman, Z and others},
  journal={Advances in Space Research},
  year={2025},
  publisher={Elsevier}
}

@article{kovacevic2020two,
  title={Two-dimensional correlation analysis of periodicity in active galactic nuclei time series},
  author={Kova{\v{c}}evi{\'c}, Andjelka B and Popovi{\'c}, Luka {\v{C}} and Ili{\'c}, Dragana},
  journal={Open astronomy},
  volume={29},
  number={1},
  pages={51--55},
  year={2020},
  publisher={De Gruyter}
}

@article{gil2010wings,
  title={Wings: Intelligent workflow-based design of computational experiments},
  author={Gil, Yolanda and Ratnakar, Varun and Kim, Jihie and Gonzalez-Calero, Pedro and Groth, Paul and Moody, Joshua and Deelman, Ewa},
  journal={IEEE Intelligent Systems},
  volume={26},
  number={1},
  pages={62--72},
  year={2010},
  publisher={IEEE}
}

@book{pruett2007yahoo,
  title={Yahoo! pipes},
  author={Pruett, Mark},
  year={2007},
  publisher={O'Reilly}
}

@article{ztf2019,
author = {Graham, Matthew and Kulkarni, S. and Bellm, Eric and Adams, Scott and Barbarino, Cristina and Blagorodnova, N. and Bodewits, Dennis and Bolin, Bryce and Brady, Patrick and Cenko, Stephen and Chang, Chan-Kao and Coughlin, Michael and De, Kishalay and Eadie, Gwendolyn and Farnham, Tony and Feindt, Ulrich and Franckowiak, Anna and Fremling, Christoffer and Gezari, Suvi and Zolkower, Jeffry},
year = {2019},
month = {07},
pages = {078001},
title = {The Zwicky Transient Facility: Science Objectives},
volume = {131},
journal = {Publications of the Astronomical Society of the Pacific},
doi = {10.1088/1538-3873/ab006c}
}

@ARTICLE{ive19,
       author = {{Ivezi{\'c}}, {\v{Z}}eljko and {Kahn}, Steven M. and {Tyson}, J. Anthony and {Abel}, Bob and {Acosta}, Emily and {Allsman}, Robyn and {Alonso}, David and {AlSayyad}, Yusra and {Anderson}, Scott F. and {Andrew}, John and {Angel}, James Roger P. and {Angeli}, George Z. and {Ansari}, Reza and {Antilogus}, Pierre and {Araujo}, Constanza and {Armstrong}, Robert and {Arndt}, Kirk T. and {Astier}, Pierre and {Aubourg}, {\'E}ric and {Auza}, Nicole and {Axelrod}, Tim S. and {Bard}, Deborah J. and {Barr}, Jeff D. and {Barrau}, Aurelian and {Bartlett}, James G. and {Bauer}, Amanda E. and {Bauman}, Brian J. and {Baumont}, Sylvain and {Bechtol}, Ellen and {Bechtol}, Keith and {Becker}, Andrew C. and {Becla}, Jacek and {Beldica}, Cristina and {Bellavia}, Steve and {Bianco}, Federica B. and {Biswas}, Rahul and {Blanc}, Guillaume and {Blazek}, Jonathan and {Blandford}, Roger D. and {Bloom}, Josh S. and {Bogart}, Joanne and {Bond}, Tim W. and {Booth}, Michael T. and {Borgland}, Anders W. and {Borne}, Kirk and {Bosch}, James F. and {Boutigny}, Dominique and {Brackett}, Craig A. and {Bradshaw}, Andrew and {Brandt}, William Nielsen and {Brown}, Michael E. and {Bullock}, James S. and {Burchat}, Patricia and {Burke}, David L. and {Cagnoli}, Gianpietro and {Calabrese}, Daniel and {Callahan}, Shawn and {Callen}, Alice L. and {Carlin}, Jeffrey L. and {Carlson}, Erin L. and {Chandrasekharan}, Srinivasan and {Charles-Emerson}, Glenaver and {Chesley}, Steve and {Cheu}, Elliott C. and {Chiang}, Hsin-Fang and {Chiang}, James and {Chirino}, Carol and {Chow}, Derek and {Ciardi}, David R. and {Claver}, Charles F. and {Cohen-Tanugi}, Johann and {Cockrum}, Joseph J. and {Coles}, Rebecca and {Connolly}, Andrew J. and {Cook}, Kem H. and {Cooray}, Asantha and {Covey}, Kevin R. and {Cribbs}, Chris and {Cui}, Wei and {Cutri}, Roc and {Daly}, Philip N. and {Daniel}, Scott F. and {Daruich}, Felipe and {Daubard}, Guillaume and {Daues}, Greg and {Dawson}, William and {Delgado}, Francisco and {Dellapenna}, Alfred and {de Peyster}, Robert and {de Val-Borro}, Miguel and {Digel}, Seth W. and {Doherty}, Peter and {Dubois}, Richard and {Dubois-Felsmann}, Gregory P. and {Durech}, Josef and {Economou}, Frossie and {Eifler}, Tim and {Eracleous}, Michael and {Emmons}, Benjamin L. and {Fausti Neto}, Angelo and {Ferguson}, Henry and {Figueroa}, Enrique and {Fisher-Levine}, Merlin and {Focke}, Warren and {Foss}, Michael D. and {Frank}, James and {Freemon}, Michael D. and {Gangler}, Emmanuel and {Gawiser}, Eric and {Geary}, John C. and {Gee}, Perry and {Geha}, Marla and {Gessner}, Charles J.~B. and {Gibson}, Robert R. and {Gilmore}, D. Kirk and {Glanzman}, Thomas and {Glick}, William and {Goldina}, Tatiana and {Goldstein}, Daniel A. and {Goodenow}, Iain and {Graham}, Melissa L. and {Gressler}, William J. and {Gris}, Philippe and {Guy}, Leanne P. and {Guyonnet}, Augustin and {Haller}, Gunther and {Harris}, Ron and {Hascall}, Patrick A. and {Haupt}, Justine and {Hernandez}, Fabio and {Herrmann}, Sven and {Hileman}, Edward and {Hoblitt}, Joshua and {Hodgson}, John A. and {Hogan}, Craig and {Howard}, James D. and {Huang}, Dajun and {Huffer}, Michael E. and {Ingraham}, Patrick and {Innes}, Walter R. and {Jacoby}, Suzanne H. and {Jain}, Bhuvnesh and {Jammes}, Fabrice and {Jee}, M. James and {Jenness}, Tim and {Jernigan}, Garrett and {Jevremovi{\'c}}, Darko and {Johns}, Kenneth and {Johnson}, Anthony S. and {Johnson}, Margaret W.~G. and {Jones}, R. Lynne and {Juramy-Gilles}, Claire and {Juri{\'c}}, Mario and {Kalirai}, Jason S. and {Kallivayalil}, Nitya J. and {Kalmbach}, Bryce and {Kantor}, Jeffrey P. and {Karst}, Pierre and {Kasliwal}, Mansi M. and {Kelly}, Heather and {Kessler}, Richard and {Kinnison}, Veronica and {Kirkby}, David and {Knox}, Lloyd and {Kotov}, Ivan V. and {Krabbendam}, Victor L. and {Krughoff}, K. Simon and {Kub{\'a}nek}, Petr and {Kuczewski}, John and {Kulkarni}, Shri and {Ku}, John and {Kurita}, Nadine R. and {Lage}, Craig S. and {Lambert}, Ron and {Lange}, Travis and {Langton}, J. Brian and {Le Guillou}, Laurent and {Levine}, Deborah and {Liang}, Ming and {Lim}, Kian-Tat and {Lintott}, Chris J. and {Long}, Kevin E. and {Lopez}, Margaux and {Lotz}, Paul J. and {Lupton}, Robert H. and {Lust}, Nate B. and {MacArthur}, Lauren A. and {Mahabal}, Ashish and {Mandelbaum}, Rachel and {Markiewicz}, Thomas W. and {Marsh}, Darren S. and {Marshall}, Philip J. and {Marshall}, Stuart and {May}, Morgan and {McKercher}, Robert and {McQueen}, Michelle and {Meyers}, Joshua and {Migliore}, Myriam and {Miller}, Michelle and {Mills}, David J.},
        title = "{LSST: From Science Drivers to Reference Design and Anticipated Data Products}",
      journal = {\apj},
         year = 2019,
        month = mar,
       volume = {873},
       number = {2},
          eid = {111},
        pages = {111},
          doi = {10.3847/1538-4357/ab042c},
archivePrefix = {arXiv},
       eprint = {0805.2366},
 primaryClass = {astro-ph},
       adsurl = {https://ui.adsabs.harvard.edu/abs/2019ApJ...873..111I}
}

@misc{ale25,
author = {Alertissimo},
  title        = {Alertissimo},
  year         = {2025},
  howpublished = {\url{https://github.com/sambolino/alertissimo}},
  note         = {Accessed: 2025-10-29}
}

\clearpage

\appendix{DSL for Transients - example code}
\label{appendix}

Here is a sample code of DSL for Transients for supermassive binary black hole (SMBBH) detection, based on the use case presented on the online meeting of the LSST-TVS collaboration. The use case was proposed by prof dr Andjelka Kova\v{c}evi\'{c} based on longstanding research of AGN and SMBBH variability (\citet{kovacevic2020two}, \citet{komossa2025extremes}). The meeting took place on the Feb 21st 2025. Comments include original definition of the use case from the discussion.

\begin{lstlisting}[style=myListingStyle]
# Fink+ALeRCE+Lasair+ANTARES agree on alert 
# ('required' argument is added for showcase)
confirm object_id="ZTF25aazqavg" brokers=[fink, alerce, lasair, antares] required=3
# ALeRCE retrieves ZTF curves of alerted object
lightcurve broker=alerce survey="ztf"
# Lasair retrieves historical light curves from Pan-STARR of 
# alerted objects, and crossmatches with IR, R, X catalogues.  
# Helps identify multi-wavelength periodicity.
crossmatch broker=lasair catalog="panstarrs" filters=["ir", "radio", "xray"]
# ANTARES crossmatches alerted object with eROSITA etc for 
# further confirmation of multiwavelength of periodicity origin.
crossmatch broker=antares catalog="erosita"
# Lasair provides a Kafka-based alert stream that updates in 
# real-time whenever a new observation is made for a 
# monitored object.
monitor broker=lasair stream="kafka"
classify method="periodicity_detection"
notify team
store db
\end{lstlisting}

\end{document}